\documentclass[%
reprint,
 amsmath,amssymb,
 aps,showkeys,
prl
]{revtex4-1}

\usepackage{graphics}
\usepackage{graphicx}% Include figure files
\usepackage{dcolumn}% Align table columns on decimal point
\usepackage{bm}% bold math
\usepackage{hyperref}% add hypertext capabilities
\usepackage{placeins}

\usepackage{colortbl}

\newcommand\erfc{\mbox{erfc}}

\begin{document}
\title{Median and Mode in First Passage under Restart}
\date\today

\author{Sergey Belan}
\email{sergb27@yandex.ru}
%\author{Mehran Kardar}
\affiliation{Massachusetts Institute of Technology, Department of Physics, Cambridge,
Massachusetts 02139, USA}

 \begin{abstract}
Restart -- interrupting a stochastic process followed by a new start -- is known to improve 
the mean time to its completion, and the general conditions under which such an improvement is achieved are now well understood. 
Here, we explore how restart affects other important metrics of first-passage phenomena, namely the median and the mode of the first-passage time distribution. 
Our analysis provides a general criterion for when  restart lowers the median time, and demonstrates that restarting is always helpful in reducing the mode.
%Additionally, we propose a simple non-uniform restart strategy that allows to optimize the mean and the median first-passage times, regardless of the characteristic time scales of the underlying process. 
Additionally, we show that simple non-uniform restart strategies allow to optimize the mean and the median first-passage times, regardless of the characteristic time scales of the underlying process. 
These findings are illustrated with the canonical example of a diffusive search with resetting.
 \end{abstract}

\maketitle

The {\it mean first-passage time} is widely used to quantify performance in diverse applications ranging from randomized search algorithms to kinetics of chemical reactions~\cite{Redner_2001}.
Remarkably, this metrics can be significantly improved by implementation of \underline{restart}, i.e. by  interrupting the first-passage process just to start it anew. 
Speed-up by restart was first noticed  in computer science more than two decades ago~\cite{Luby_1993}, with a new wave of current interest  triggered by the seminal work of Evans and Majumdar~\cite{EM_2011} who demonstrated that stochastic resetting hastens  diffusive search.
More recently, the development of a general renewal approach has provided a unified and model independent treatment of first-passage under restart~\cite{Rotbart_2015, Reuveni_PRL_2016,Reuveni_PRL_2017}. 
In particular, it has furnished a simple criterion for when restart helps to lower the expected completion time of first-passage processes, and revealed universality in the behavior of the optimally restarted processes~\cite{Reuveni_PRL_2016}.

\begin{figure}
 \includegraphics[scale=.6]{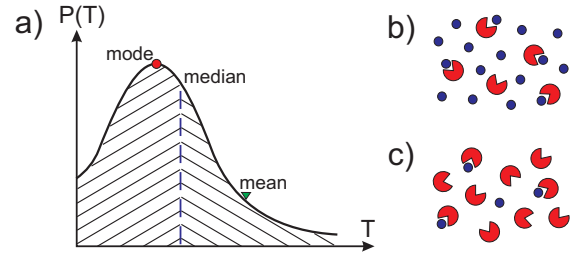}
\caption{(a): The mean $\langle T\rangle$, median $T_{1/2}$ and mode $T_m$ of some first-passage-time probability distribution $P(T)$. (b): Enzymatic reaction with an excess of substrate (schematically shown in blue). The reaction speed is determined by the mean turnover time. (c) Enzymatic reaction with excess  enzyme (schematically shown in red); the relevant time scale of the reaction is the median turnover time.}
\label{pic:setup}
 \end{figure}

While the mean completion time plays a central role for some applications, in many other settings it does not capture the relevant time scale of the task, and other metrics may be more appropriate to quantify performance. 
More specifically, there are cases when the {\it median} time should be used instead of the mean.
For example, in an enzymatic reaction with an excess of substrate molecules (see Fig.~\ref{pic:setup}b), the time taken for a significant change in concentration of substrate is large compared with the expected catalysis time, and thus 
 the later is a more natural measure of how fast the reaction proceeds.
 Indeed, in the limit of large substrate concentration, the number of substrate molecules converted to products in a unit volume per second is proportional to inverse mean catalysis time~\citep{Reuveni_2014,Reuveni_2018}. 
However, in the opposite case when concentration of substrate molecules is low compared to that of the enzyme molecules (Fig.~\ref{pic:setup}c), one deals with the highly non-stationary situation characterized by fast substrate depletion~\cite{Segel_1989,Borghans_1996,Schnell_2000}, and the mean catalysis time may be not informative.
The natural metric of the reaction speed in such a non-stationary situation is given by the median turnover time -- the time required to convert half of the initial amount of substrate into product. 
Also, in the contexts of randomized search algorithms and Internet tasks, one may be interested in the typical behavior captured by the median completion time rather than in the average values which may be dominated by rare but extreme runs~\cite{Luby_1993,Moorsel_2004,Wu_2006,Lorenz_2016}.

Here, we analyze the effect of restart on the median time of a generic first-passage process. 
Namely, our analysis provides a condition for when introduction of a small restart rate reduces the median first-passage time, or more generally, a given \textit{quantile} of the first-passage-time density (FPTD).
Using diffusive search as illustrative example, we show that similarly to the mean first-passage time, the median can be optimized by a careful choice of the restart rate.
Since the optimal restart rate is determined by the time scale of the first-passage process, which is often unknown a priory, we also explore restart protocols  whose performance is weakly sensitive to such details. 
In addition, we describe the effect of restart on the {\it mode} of the first-passage time distribution, i.e. on the value of the process completion time which occurs most often.
It turns out that in contrast to the mean and median, which may increase or decrease in
response to introduction of restart (depending on  process details), this metric cannot be increased by restart.

Consider a first-passage time process characterised by the random completion time $T$, with the FPTD $P(T)$. 
By definition, the $q$th quantile is the time $T_q$ such that the process has the probability $0<q<1$ to finish before $T_q$.
Say, for $q=1/2$,  $T_{1/2}$  represents the median first-passage time. 
Clearly, this quantity obeys the following integral equation
\begin{equation}
\label{quantile}
\int\limits_0^{T_q} dT P(T)=q.
\end{equation}
Now let us assume that the process becomes subject to stochastic restart at the infinitesimally small rate $\delta r\to\,+0$.
How does this affect the $q$th quantile of this process?
In the presence of restart, Eq.~(\ref{quantile}) takes the form $\int\limits_0^{T_q+\delta T_q} dT P_{\delta r}(T)=q$, 
where $\delta T_q$ represents the change in the $q$th quantile and $P_{\delta r}(T)$ is the FPTD modified by the restart.
It is straightforward to show that
\begin{equation}
\label{quantile03}
\delta T_q=-\frac{1}{P(T_q)}\int\limits_0^{T_q} dT [P_{\delta r}(T)-P(T)]\,.
\end{equation}

To proceed we need to know the difference $P_{\delta r}(T)-P(T)$  which determines the response of the FPTD to the introduction of rare Poisson restarts.
As shown in Refs.~\cite{Rotbart_2015, Reuveni_PRL_2016}, the Laplace transform of the FPTD $\tilde P_r(s)=\int_0^\infty dT e^{-sT}P_r(T)$ of any stochastic process under constant restart rate $r$ is given by 
\begin{equation}
\tilde{P}_r(s)=\frac{\tilde{P}(s+r)}{\frac{s}{s+r}+\frac{r}{s+r}\tilde{P}(s+r)}\,.
\end{equation}
In the limit of  small restart rate $\delta r$, this equation yields 
\begin{equation}
\tilde{P}_{\delta r}(s)\approx\tilde{P}(s)+\left(\frac{\partial\tilde{P}(s)}{\partial s}-\frac{1}{s}\tilde{P}(s)^2+\frac{1}{s}\tilde{P}(s)\right)\delta r\,.
\end{equation}
Using the identities
\begin{equation}
\int_0^\infty dT P(T)Te^{-sT}=-\partial_s\tilde{P}(s),
\end{equation}
\begin{equation}
\int_0^\infty dT e^{-sT}\int_0^Tdt P(t)=\frac{1}{s}\tilde{P}(s),
\end{equation}
\begin{equation}
\int_0^\infty dT e^{-sT}\int_0^Tdt\int_0^{t}d\tau P(\tau)P(t-\tau)=\frac{1}{s}\tilde{P}^2(s),
\end{equation}
%which can be straightforwardly checked by direct calculations, 
we perform the inverse Laplace transform to obtain  to the linear-order  in $\delta r$,
\begin{eqnarray}
\!\! P_{\delta r}(T)\!- P(T)\!\approx\left(\int\limits_T^\infty dt(P_2(t)-P(t))\!-\!
\label{Laplace02}
P(T)T\right)\!\delta r,
\end{eqnarray}
where $P_2(t)\equiv\langle\delta(T_1+T_2-t)\rangle_{T_1,T_2}=\int_0^t d\tau P(\tau)P(t-\tau)$ is the probability distribution for the sum of two independent variables $T_1$ and $T_2$ sampled from $P(T)$.
Next, substitution of Eq.~(\ref{Laplace02}) into Eq.~(\ref{quantile03}) gives 
\begin{equation}
\label{Laplace002}
\delta T_q=-\frac{\delta r}{P(T_q)}\int\limits_0^{T_q} dT \left(\int\limits_T^\infty dt[P_2(t)-P(t)]-P(T)T\right).
\end{equation} 

%\begin{figure}
% \includegraphics[scale=.265]{deadline}
%\caption{The deadline meeting probability $p_d$ as a function of the rate $r$ of Poisson restart for one-dimensional diffusive search of immobile target with $D=1$ and $L=1$. The restart is helpful provided a sufficiently large time margin $T_{d}>T_{d_0}\approx 0.7439 L^2/D$.}
%  \label{pic:p_d}
% \end{figure}
 
 \begin{figure}
 \includegraphics[scale=.28]{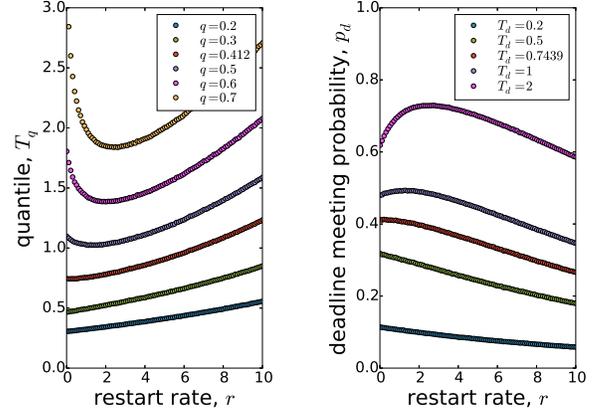}
\caption{{\it (Left)} The $q$th quantile, $T_q$, of the search time distribution as a function of the rate $r$, for Poisson restarts of a one-dimensional diffusive search for an immobile target with $D=1$ and $L=1$.  We see that, in agreement with theoretical predictions extracted from numerical analysis of Eq.~(\ref{median_criterion}), restart is helpful for $q>q_0\approx0.412$. 
{\it (Right)} The deadline meeting probability, $p_d$, as a function of the rate $r$ of Poisson restarts for a one-dimensional diffusive search for an immobile target with $D=1$ and $L=1$. Restart is helpful for a sufficiently large time margin $T_{d}>T_{d_0}\approx 0.7439 L^2/D$. The numerical data are based on  statistics including $10^6$ independent runs.}
  \label{pic:T_q}
 \end{figure}
 
From Eq.~(\ref{Laplace002}), we may conclude that a general criterion of when restart reduces the $q$th quantile (i.e. $\delta T_{q}<0$) is provided by the  inequality 
\begin{equation}
\label{median_criterion}
\int\limits_{0}^{T_q}dT \left(P(T)T+\int\limits_{0}^{T}dt (P_2(t)-P(t))\right)<0. 
\end{equation}

It is worth noting that the above analysis is also relevant to the `deadline meeting problem'~\cite{Moorsel_2004,Wu_2006, Lorenz_2016,Belan_2018}.
The probability that a first-passage process having the FPTD $P(T)$ will finish, before the prescribed deadline $T_d$ has passed, is determined by $p_d=\int\limits_0^{T_{d}} P(T)dT$.
The variational calculus based on the Eq.~(\ref{Laplace02}) shows that, when the process is subjected to a small restart rate $\delta r$, the deadline meeting probability obtains a correction $\delta p_d=\left(\int_{0}^{T_d}dT \int_{0}^{T}dt [P(t)-P_2(t)]-\int_0^{T_d}dTP(T)T\right)\delta r$.
Therefore, restart helps to increase the chance to meet deadline whenever 
$\int_0^{T_d}dT\left(P(T)T+\int_{0}^{T}dt (P_2(t)-P(t))\right)<0$.
%Eq.~(\ref{median_criterion}), with $T_d$ replacing $T_q$, is satisfied.
%\com{Check sign, read like Eq.~(\ref{median_criterion}), with $T_d$ replacing $T_q$.}
 
 \begin{figure}
 \includegraphics[scale=.35]{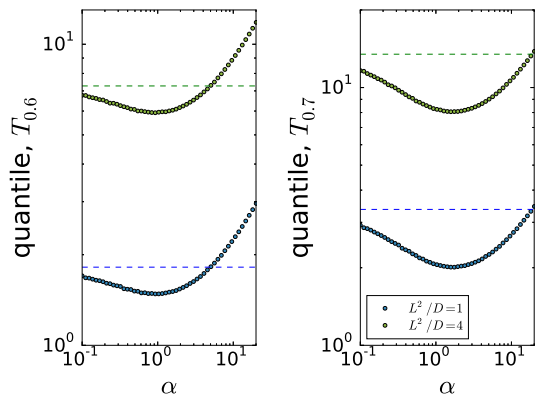}
\caption{Dependence of the quantiles of the search time distribution 
on  the dimensionless parameter  $\alpha$ of the stochastic scale-free restart protocol,
for different values of the diffusive time scale $L^2/D$. The dashed lines represent the unperturbed quantiles in the absence of restart. In numerical simulations we set a short time cut-off $\tau=10^{-3}$ to regularize the divergence of the restart rate at $t=0$, and collected statistics from $10^6$ independent realizations of the search process.}
  \label{pic:geometric quantiles}
 \end{figure}
 
Given the FPTD of any process of interest, one can then readily check if the inequality in Eq.~(\ref{median_criterion}) is fulfilled. 
For the sake of illustration, let us consider a one-dimensional Brownian motion in 
search of an immobile (absorbing) target. 
In this case, the first-passage time density is given by the Levy-Smirnov distribution: $P(T)=\frac{L}{\sqrt{4\pi DT^3}}\exp(-\frac{L^2}{4DT})$,
where $L$ is the initial distance to the target, and $D$ is the diffusion coefficient. 
It is straightforward to verify that the $q$-th quantile of this probability density is $T_q=\frac{L^2}{4D [\erfc^{-1}(q)]^2}$.
Substituting these expressions for $P(T)$ and $T_q$ into Eq.~(\ref{median_criterion}), and performing integration numerically, we find that the inequality is satisfied for $q>q_0$, where $q_0\approx 0.4123$.
Thus, the introduction of restart, which occasionally returns the particle to its initial position, reduces the median completion time $T_{1/2}$ of diffusive search. 
Stochastic simulations indicate that the median time attains a minimum at the optimal restart rate $r_0\approx 1.5D/L^2$ which is smaller than the rate $r^\ast\approx 2.5 D/L^2$ minimizing the mean search time~\cite{EM_2011}. 
Clearly, since restart works by avoiding the tail of the FPTD, it becomes more potent for larger $q$, as visible in the left panel of Fig.~\ref{pic:T_q}.
We also note in passing that restart decreases the deadline meeting probability of diffusive search for sufficiently short deadlines $T_d<T_{d_0}$ while increasing it for $T_d>T_{d_0}$, where $T_{d_0}\approx 0.7439 L^2/D$ (see the right panel in Fig.~\ref{pic:T_q}).
This is in accord with the analysis reported in Ref.~\cite{Belan_2018}, where restart is shown to increase the chance of finding a target in the presence of sufficiently small mortality rate, while reducing this chance if mortality rate is large.

\begin{figure}
 \includegraphics[scale=.35]{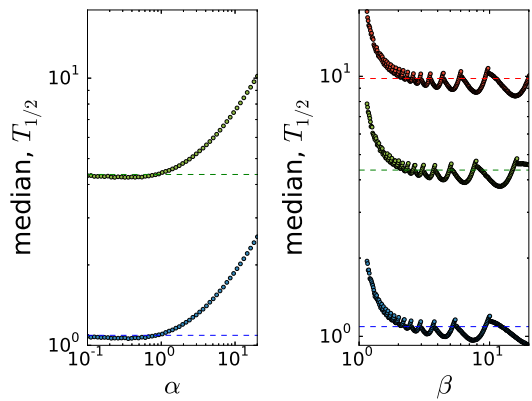}
\caption{The median $T_{1/2}$ of the search time distribution in the presence of stochastic scale-free {\it (left)}, and the geometric {\it (right)} restart protocols, for different values of the diffusive time scale $L^2/D$. The dashed lines represent  median search times in the absence of restart. }
  \label{pic:median g VS sf}
 \end{figure}
 
 Clearly, to be effective restart requires prior knowledge of the characteristic time-scale of the underlying process. 
In the case of the median search time for diffusion, the relevant rates are measured
with respect to the (inverse) diffusive time scale $\tau_{dif}=L^2/D$.
Restart rates chosen without taking this characteristic time into account may well
lead to performance that is worse than without any restart. 
Previously, a similar challenge motivated the development of the restart strategies improving the mean first-passage time without   
introducing any time scale~\cite{Luby_1993,Kusmierz_2018}.
The strategy proposed in Ref.~\cite{Kusmierz_2018} is make restarts
separated by random scale-free time intervals.
We numerically investigated the effect of such stochastic scale-free restarts 
on the quantiles of the diffusive search. 
Specifically, we implement a non-uniform restart rate which is inversely proportional to the time elapsed since the start of the process,
i.e. $r(t)=\frac{\alpha}{t}$, where $\alpha$ is a dimensionless constant. 
Figure~\ref{pic:geometric quantiles} demonstrates that such a non-uniform restart protocol allows to minimize the large-$q$ quantiles of the FPTD without being sensitive to the parameters of the problem. 
Indeed, the quantiles attain extrema at the optimal values $\alpha^\ast_q$ which do not depend on the diffusive time $\tau_{dif}$ 
in contrast to the optimal rate of uniform restart which scales proportionally to $\tau_{dif}$.

\begin{figure}
 \includegraphics[scale=.35]{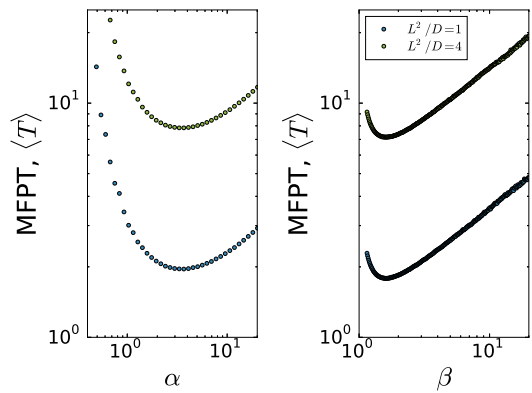}
\caption{The mean search time $\langle T \rangle$ in the presence of stochastic scale-free {\it (left)} and the geometric {\it (right)} restart protocols  for different values of the diffusive time scale $L^2/D$.}
  \label{pic: MFPT geometric VS scalefree}
 \end{figure}

We also propose and explore a non-uniform deterministic restart protocol with restart times chosen from a geometric sequence, i.e. with the $n$th restart event at $t_n=\tau \beta^{n-1}$, where $\tau\ll \tau_{dif}$ is a microscopic cut-off, and $\beta$ is a dimensionless constant. 
In a long run, the time interval between successive restarts approaches  the  elapsed time since the start of the process, so that there is no characteristic restart frequency.
As depicted in Fig.~\ref{pic:median g VS sf}b, we find that $T_q$ obtains an oscillatory
dependence on $\beta$, which renders practical implementation of this strategy 
of quantile optimization problematic.
Note, however, that the geometric restart protocol is actually quite efficient for 
reducing the mean search time, see Fig.~\ref{pic: MFPT geometric VS scalefree}.
What is more, the performance of the geometric restart in terms of the MFPT is better than that of the above mentioned stochastic scale-free strategy. 
The former achieves a minimal mean search time of $\langle T_{\beta^\ast}\rangle\approx 1.8 \tau_{dif}$ at the optimal value $\beta_\ast\approx 1.6$ that is not sensitive to the diffusive time, while the latter gives $\langle T_{\alpha^\ast}\rangle\approx 1.97 \tau_{dif}$ at $\alpha_\ast\approx 3.5$~\cite{Kusmierz_2018}.

Finally let us discuss another interesting metric of  first-passage processes: 
the mode of the FPTD, i.e. the time $T_m$ at which the probability distribution $P(T)$ 
takes its maximum value.
In other words, $T_m$  is the value of the  completion time $T$ that occurs most often,
which must thus satisfy $\frac{dP(T)}{dT}|_{T=T_m}=0$, together with  $\frac{d^2P(T)}{dT^2}|_{T=T_m}<0$.
In the presence of restarts at the small rate $\delta r$, we have $\frac{dP_{\delta r}(T)}{dT}|_{T=T_m+\delta T_m}=0$, leading to
\begin{equation} 
\delta T_m=-\frac{1}{\frac{d^2P(T)}{dT^2}|_{T=T_m}}\frac{d}{dT}[P_{\delta r}(T)-P(T)]|_{T=T_m}\,.
\end{equation}
Next, using Eq.~(\ref{Laplace02}) one obtains 
\begin{equation}
\label{mode}
\delta T_m=\frac{\delta r}{\frac{d^2P(T)}{dT^2}|_{T=T_m}}\int\limits_0^{T_m}d\tau P(\tau)P(T_m-\tau).
\end{equation}
Since $\frac{d^2P(T)}{dT^2}|_{T=T_m}<0$ and $\int_0^{T_m}d\tau P(\tau)P(T_m-\tau)=P_2(T_m)\geq 0$ , we immediately find that $\delta T_m\leq 0$.
Thus, introduction of restart decreases or leave unchanged the mode of any FPTD.
%The mode remains the same only if it is initially equal to zero. 
Although this conclusion is based on the assumption of the infinitesimally small restart rate, the same result remains valid for any FPTD.
Indeed, based to the splitting rule of the Poisson process, we can safely assume that $P(T)$ in the above formulas represents the FPTD of the process that is already subject to restart at some non-vanishing rate.
Then Eq.~(\ref{mode}) indicates that the mode of this process is a non-increasing function of the restart rate $r$.

In conclusion,  the effectiveness of restarts in reducing the
{\it mean first-passage time} has been demonstrated in a number of studies ~\cite{EM_2011,Evans_2011,Whitehouse_2013,Evans_2014,Kusmierz_2014,Kusmierz_2015,Pal_2016,Rotbart_2015,Reuveni_PRL_2016,
Reuveni_PRL_2017,Kusmierz_2018,Eule_2016,Nagar_2016,Pal_2019a,Pal_2019b,Campos_2019,Giuggioli_2018,Kusmierz_2019b,Masoliver_2019,Evans_2019a,Evans_2018a}.
However, less has been known about how restart affects other characteristic time 
metrics of the first-passage completion.
To fill this gap,  we have explored the advantages of restarting to optimization of the median and mode of a generic first-passage-time density.

The ubiquity of restarts in natural and artificial systems encourages us to think that ideas presented here will find diverse applications. 
Say, in the context of enzymatic reactions, restarts  correspond to unbinding of substrate from enzyme prior to completion of the catalytic step.
Indeed, a previous analysis in the limit of large substrate concentration has showed that increase of the substrate unbinding rate can accelerate the reaction~\cite{Reuveni_2014}.   
Our results suggest that a similar effect can potentially be achieved in the opposite limit of large enzyme concentration when the reaction speed is determined by the substrate ``half-life," as mentioned in the introductory part of this work.

A range of issues call for further theoretical investigation.
The geometric restart protocol proposed here, and the non-uniform scale-free stochastic restart explored in~\cite{Kusmierz_2018}, allow us to hypothesize the existence of still undiscovered family of non-uniform restart strategies whose performance is weakly sensitive to underlying details.
An intersting open question is if there is a single universally optimal strategy in this family, 
i.e. a strategy that achieves the best performance  for any  first-passage process. 
A recent study revealed that in the class of uniform restarts this property is exhibited by the deterministic (regular) restart~\cite{Reuveni_PRL_2017}.
This allows us to speculate that, being both deterministic and scale-free, the geometric restart may possess important optimal features.
Another unsettled issue is if it is possible to derive a rigorous bound on the performance of the geometric restart protocol similarly to what was previously obtained for the Luby's universal strategy in computer science applications~\cite{Luby_1993}.

\begin{acknowledgments}
S. B. gratefully acknowledges support from the James S. McDonnell Foundation via its postdoctoral fellowship in studying
complex systems. S.B. acknowledges support from NSF through grant DMR-1708280.
S. B. would like to thank Mehran Kardar for reading the paper and for providing comments.
\end{acknowledgments}

{}

\end{document}